\documentclass[%
 aip,
rsi,%
 amsmath,amssymb,
 reprint,%
]{revtex4-1}

\usepackage{graphicx}
\usepackage{dcolumn}
\usepackage{bm}

\begin{document}

\title[]{Demonstration of a compact linear accelerator}
\author{P.A. Seidl}
 \email{paseidl.lbl.gov}
\author{A. Persaud}
\author{W. Ghiorso}
\author{Q. Ji}
\author{W.L. Waldron}
\affiliation{ 
E.O. Lawrence Berkeley National Laboratory, 1 Cyclotron Road, Berkeley, CA 94720, USA}%

\author{A. Lal}
\author{K.B. Vinayakumar}
\affiliation{%
SonicMEMS Laboratory, Cornell University, Ithaca, NY 14853, USA
}%
\author{T. Schenkel}
\affiliation{ 
E.O. Lawrence Berkeley National Laboratory, 1 Cyclotron Road, Berkeley, CA 94720, USA}%

\begin{abstract}
Recently, we presented a new approach for a compact radio-frequency (RF) accelerator structure and demonstrated the functionality of the individual components: acceleration units and focusing elements. In this paper, we combine these units to form a working accelerator structure including a matching section between the ion source extraction grids and the RF-acceleration unit. The matching section consist of six electrostatic quadrupoles (ESQs) fabricated using 3D-printing techniques. The matching section enables us to capture twice the amount of beam and match the beam envelope to conditions for an acceleration lattice. We present data from an integrated accelerator consisting of the source, matching section, and an ESQ doublet sandwiched between two RF-acceleration units.
\end{abstract}


\maketitle
%
%
%
%
\section{Introduction}
We recently presented a new approach to enable compact RF particle accelerators for the generation of high intensity ion beams.  The concept is motivated by early research on the acceleration of many parallel beams for heavy ion driven inertial fusion energy.  This earlier work showed that for extreme beam current applications, the considerable space charge could be feasibly managed by accelerating many parallel, closely-spaced lower current beams more economically than a small number of beams in separate accelerator structures. 

Focusing and acceleration fields are limited by electrical breakdown, and for a quadrupole focusing element, a maximum operating field on the electrode will produce a stronger focusing gradient in a smaller diameter channel.  Maschke \cite{Maschke_1979,Maschke_1979b} showed that as the multiple beam array electrostatic focusing electrodes and apertures are scaled to smaller dimensions, the average current density scales favorably, and increases until alignment tolerances and vacuum pressure in the dense structure lead to excessive emittance growth and beam loss.  Experiments with a prototype multiple-beam RF accelerator demonstrated the concept for a beam-beam separation of 1.3 cm. \cite{URBANUS1989508}  
Similarly, others were motivated to address high-space charge with closely packed quadrupole focusing arrays in a common induction accelerator for heavy ion inertial fusion energy~\cite{Bangerter2013} using electrostatic and magnetic focusing quadrupoles. 

Recent developments of micro-electromechanical systems (MEMS) have demonstrated impressive fabrication tolerances at very low cost, opening the way to mm-scale densely packed beams with electrostatic focusing and RF acceleration.  Our first results with mm-scale structures are described in Refs. [\onlinecite{Persaud}] and [\onlinecite{Persaud2}].

Scaling to a large number of densely packed beams may lead to various  applications, such as mass analysis for ion implantation (few mA and up to 300 keV ion energy). \cite{Hamm12} At higher current per beam, with each beam approaching the focusing limits set by space charge repulsion and beam loading in the RF system, the compact accelerator may meet the extreme beam requirements for ion beam heating of plasmas for fusion energy applications. \cite{Sonato17}

In this paper, we focus on the challenge of the transverse matching of many round beams from the ion source to the alternating-gradient focusing channels of the accelerator.  Beams injected from ion sources are usually cylindrically symmetric in profile, and may be slightly converging or diverging depending on the optics in the injector and the beam properties such as emittance and space charge.  For injection into accelerators with alternating-gradient quadrupole focusing systems, the matched beam conditions midway between quadrupoles are generally a cylindrically symmetric spatial profile (x-y space) but the beam almost always is converging in one plane and diverging in the other.  This may be seen from the solutions to the rms envelope equation for an alternating gradient quadrupole focusing lattice \cite{Reiser1994}.  To minimize particle loss and emittance growth, a special matching section of several quadrupoles is often used to transform the beam envelope from the round conditions at the exit of the injector to matched beam conditions in the accelerator lattice.

\section{Matching Section Design}
The rms transverse envelope equations describes the rms beam size as a function of the propagation direction subject to the applied electrostatic focusing fields from the quadrupoles, the effective defocusing due to the beam emittance, and the defocusing self-field of the ion beam.  The system has four-fold symmetry and the horizontal and vertical equations are coupled via the space charge term:
\begin{equation}
a'' = \kappa a + \frac{\epsilon^2}{a^3}+\frac{Q}{a+b}
\end{equation}

and 
\begin{equation}
b'' = -\kappa b + \frac{\epsilon^2}{b^3}+\frac{Q}{a+b}
\end{equation}

\begin{figure}[ht!]
  \centering
  \includegraphics[width=0.95\linewidth]{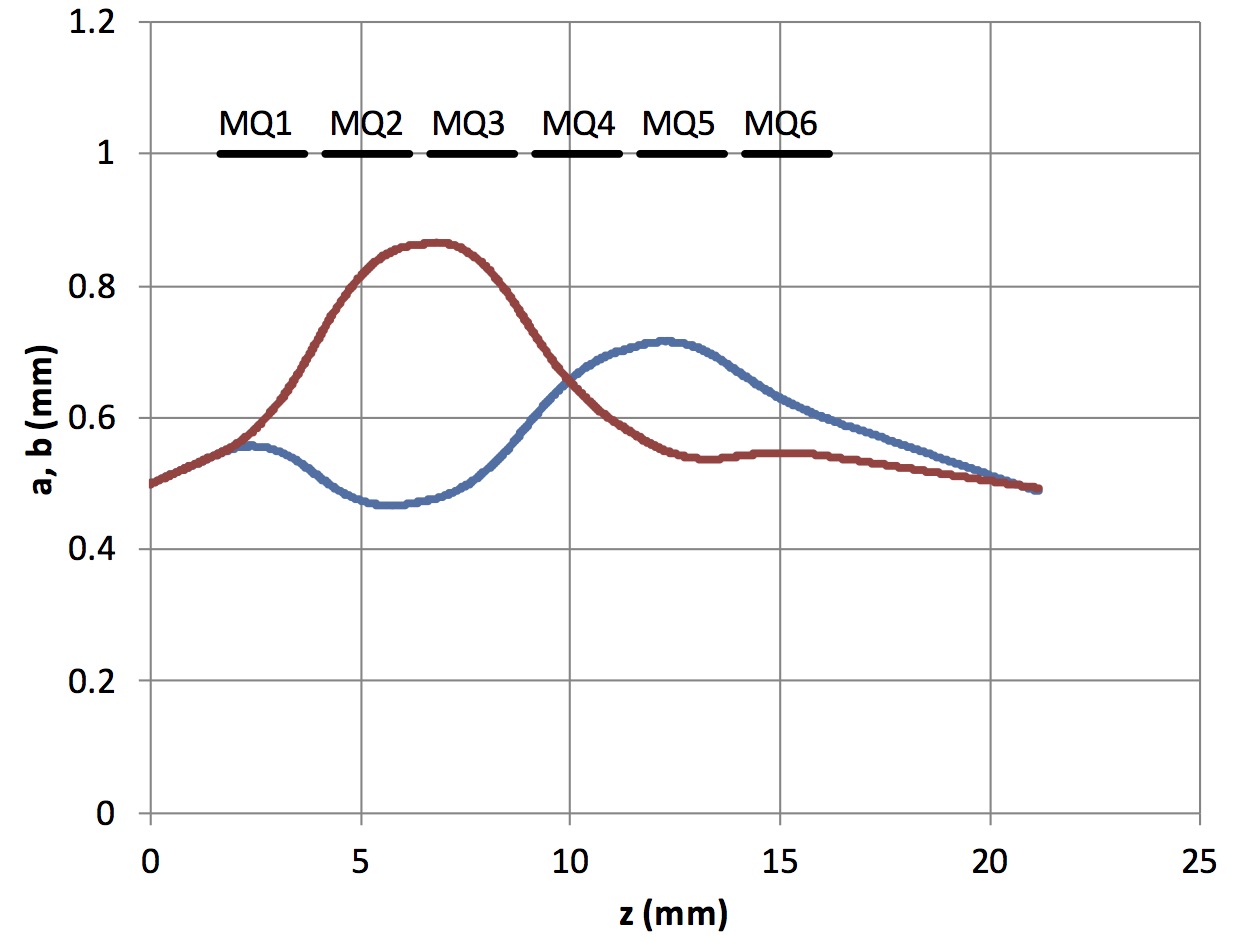}
  \caption{(Color online) An example RMS envelope solution for a single beam.  The assumed initial conditions of the beam are based on an average measurement of the beam envelope at the exit of the injector.  The quadrupole voltages are chosen to establish a converging beam condition at the entrance of the accelerator section immediately downstream of the six-quadrupole matching section. The horizontal lines indicate the length and positions of the quadrupole apertures. }
  \label{fig:match-envelope1}
\end{figure}

The derivatives are with respect to the beam propagation axis ($z$), and $a$ and $b$ are the horizontal and vertical $2 \cdot rms$ envelope coordinates of the beam envelope. The quadrupole focusing strength, $4 \cdot rms$ un-normalized emittance and dimensionless perveance of the beam are denoted by $\kappa$, $\epsilon$, and $Q$, respectively.  For non-relativistic particles

\begin{equation} 
Q = \frac{\lambda}{4 \pi \epsilon_0 V_{b}} 
\end{equation}
and 
\begin{equation}
\kappa = \frac{V}{V_{b}R^2}
\end{equation}
where $\lambda$ is the beam line charge density, $V$ is the quadrupole potential, $R$ is the radius from the aperture center to the electrode and $V_{b}$ is the beam voltage.
The beam envelope is simulated via numerical integration of the envelope equations, initialized by the beam parameters ($a_i$, $a_i'$, $b_i$ and $b_i'$) at the exit of the injector.  These parameters were measured by imaging the beam profile at two axial locations downstream of the injector (without the matching section. A minimum of four quadrupoles is needed to solve for the quadrupole focusing strengths to match the beam envelope to the periodic focusing solution of the downstream elements: $a(z) = a(z+L)$, where $L$ is the lattice period of the quadrupole focusing lattice in the accelerator and similarly for $a_i'$, $b_i$ and $b_i'$.  For these first experiments with the matching section, we chose solutions to establish a simpler, converging beam at the entrance to the accelerator.  Fig.~\ref{fig:match-envelope1} shows a solution which transforms the axi-symmetric diverging beam at the exit of the injector to a converging beam downstream of the matching section with a similar beam radius.  To achieve this, we have used a FDDFFD solution instead of an alternating gradient pattern between focusing (F) and defocusing (D) in a given plane between successive quadrupoles (FDFD\ldots). 

The input beam parameters are shown in Table~\ref{tab:simulation}, and the quadrupole voltages are all less than 600 Volts to be established in a bipolar manner.   
The solution is for an 8~keV Ar$^+$ ion beam (I = 7 $\mu$A, $\epsilon = 0.63\, \mu m$).  However, the quadrupole voltage solution is insensitive to the ion mass species and can transport a variety of ion species without extensive retuning.  

\begin{figure}[ht!]
  \centering
  \includegraphics[width=0.95\linewidth]{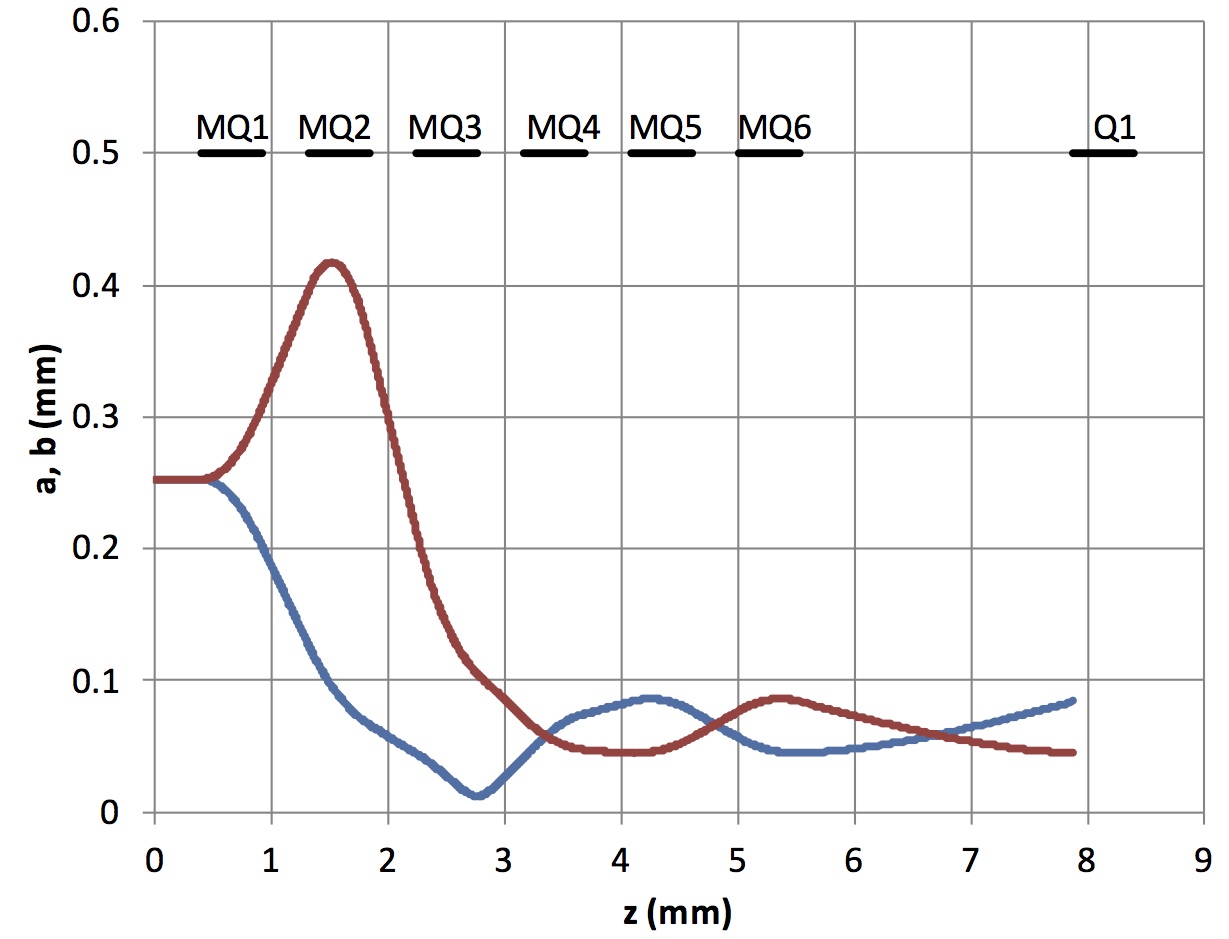}
  \caption{(Color online) Matched beam solution for a more compact structure, with 1-mm diameter apertures.}
  \label{fig:match-envelope2}
\end{figure}

An envelope solution that establishes matched conditions at the beginning of an accelerator section is shown in Fig.~\ref{fig:match-envelope2}. 

\begin{table}
\caption{\label{tab:simulation}Beam parameters and quadrupole voltages for the simulation in Fig.~\ref{fig:match-envelope1}.}
\begin{ruledtabular}
\begin{tabular}{llll}
$a_i$ (mm) &0.5 &$V_1$ (V) & 552\\
$a_i'$ (rad) &0.028 &$V_2$ (V) & -500\\
$\epsilon$ (mm) &6.3e-4 &$V_3$ (V) &-512\\
%
%
%
%
%
%
$E_{ion}$ (keV) &8.0 &$V_4$ (V) & 359\\
mass (amu) &40 &$V_5$ (V) & 352\\
 & &$V_6$ (V) & -136\\
\end{tabular}
\end{ruledtabular}
\end{table}

\section{Quadrupole Fabrication and Experimental Setup}
\begin{figure}[ht!]
  \centering
  \includegraphics[width=0.95\linewidth]{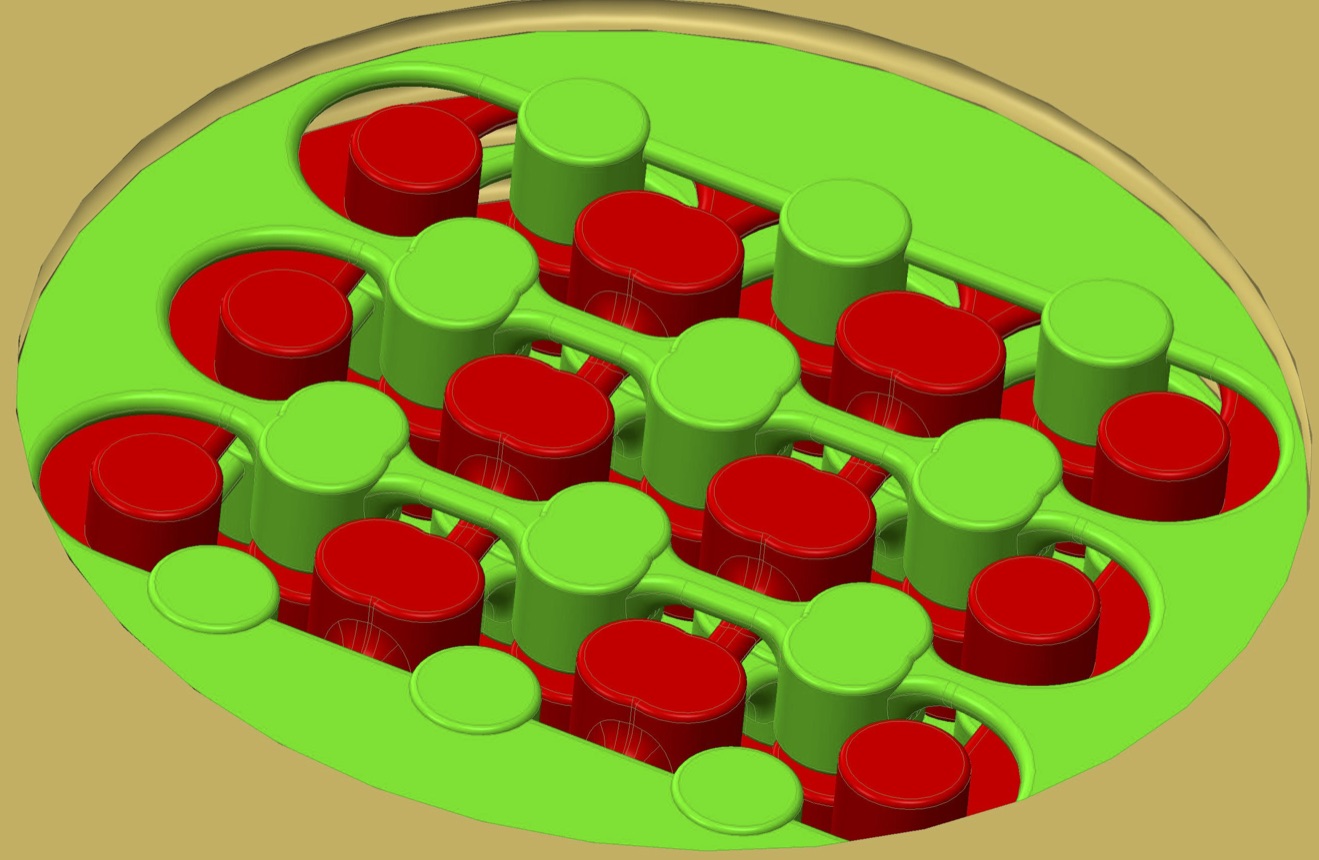}
  \caption{(Color online) The CAD model shows the cylindrical electrodes (red and green) supported by thin ribs. Each electrode part holds all the electrodes for one polarity of the nine-beam array, and is held in place by an outer plastic holder (brown). The ribs provide mechanical support and electrical connection to the DC power supplies.}
  \label{fig:quad-detail}
\end{figure}

The principal components of the mechanical assembly are the electrodes which when electrically biased define the quadrupole field pattern, and the support frames which support the electrodes and maintain overall alignment.  The electrodes are electroless nickel-coated on a monomer resin substrate \cite{protolabs}.  The substrates were manufactured with a rapid prototype, 3D-printing process, leading to the high precision (25 $\mu$m) needed to maintain the electrode shapes and spacings for good alignment and field quality while maintaining low component cost.  The electrode components are held in polyether ether ketone (PEEK) plastic support frames.  Since this is a multiple beam structure, each polarity electrode for a quadrupole array is supported by connecting ribs to enable close spacing (Fig.~\ref{fig:quad-detail}).   

\begin{figure}[ht!]
  \centering
  \includegraphics[width=0.95\linewidth]{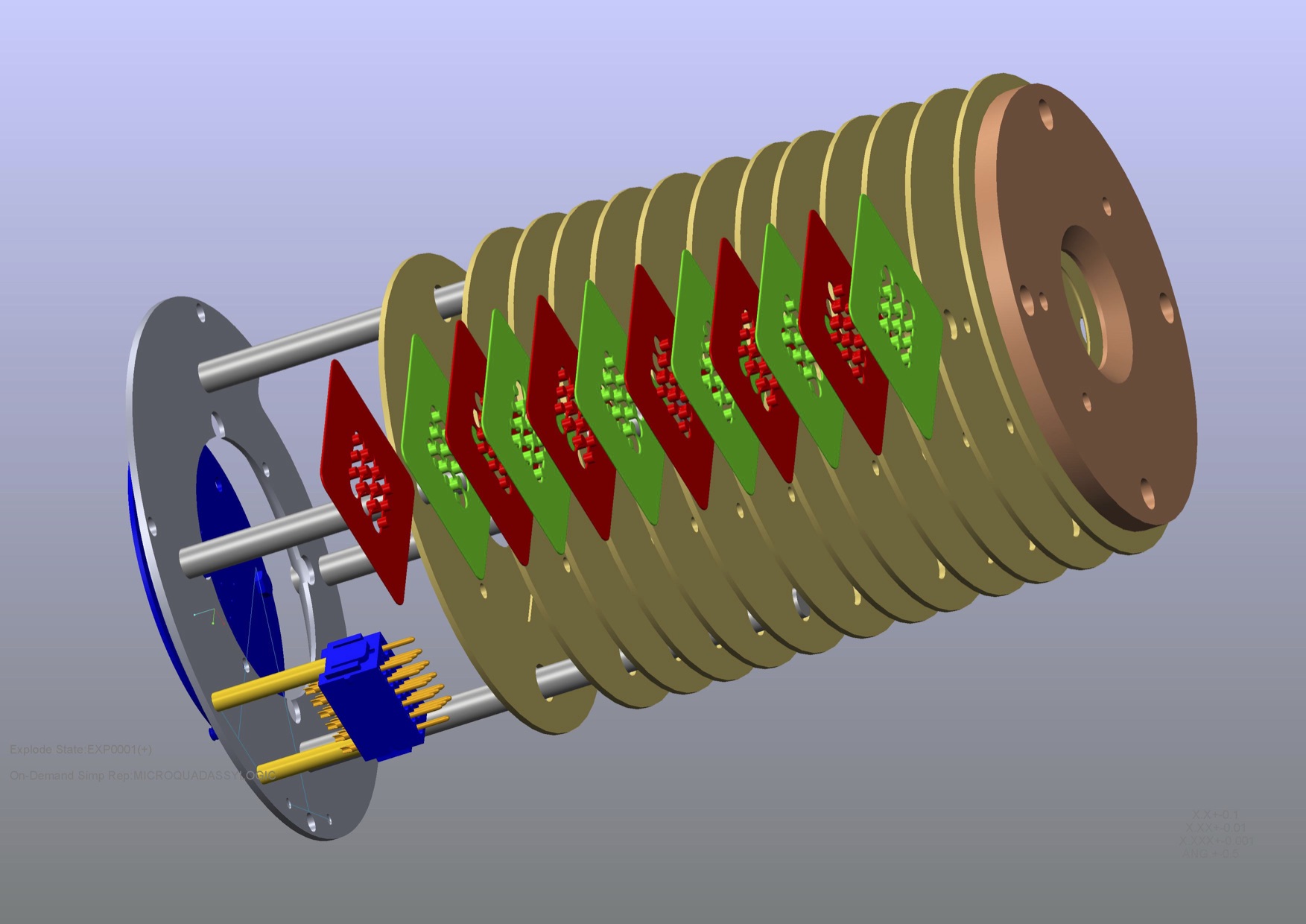}
  \caption{(Color online) The CAD model of the quadrupole assembly is shown in an exploded view for clarity.  The quadrupole electrodes (highlighted in red and green) are supported within the larger plastic (PEEK) holders. The holders are in turn held in alignment by stainless steel rods mounted to a base plate. }
  \label{fig:quad-assembly}
\end{figure}

One set of electrodes overlays the opposing polarity electrodes and they fully overlap in $z$ forming a quadrupole of length 2~mm with gaps of 0.5~mm between successive quadrupoles. Electrical continuity to the electrodes is via soldered connections to a vapor deposited gold coating on the PEEK supports, which in turn is in contact with the nickel coated quadrupole electrodes.  The nickel coating process slightly warped
the quadrupole electrode parts.  However, when held in place by the more rigid PEEK holders, the electrodes were restored to the desired coplanar orientation. A coordinate measuring survey of the assembly showed beam-center spacing (pitch between beams: 5.0 mm) errors and electrode diameter errors well within the 25 $\mu$m tolerance specifications. 
The CAD model of the assembly is shown in Fig.~\ref{fig:quad-assembly}, illustrating that a $3 \times 3$ transverse array of quadrupoles is comprised of a pair of electrode parts where one part (green) will be biased to, e.g., $+V$ for the horizontally oriented electrodes and the vertically oriented electrodes (red) are biased to $-V$. Thus, the voltages to establish a particular matching solution are common to all nine beams.

\begin{figure}[ht!]
  \centering
  \includegraphics[width=0.95\linewidth]{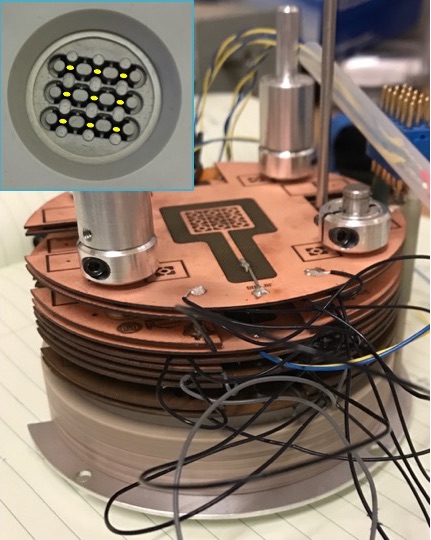}
  \caption{(Color online) The photograph of the final assembly shows the most downstream matching quadrupole (inset) within the outer support structure. The beams are represented schematically by the yellow ellipses. In the main photo, the matching section is at the bottom of the assembly, and only the plastic supports are visible.  The RF accelerator section, fabricated from laser-cut 10-cm diameter PC-boards is aligned via the common alignment rods at gaps between the elements are set to match the RF frequency and ion velocity.}
  \label{fig:matching-photo}
\end{figure}

The assembled matching section is shown in Fig.~\ref{fig:matching-photo}.  Note that there is considerable radial space available within this structure footprint to build a quadrupole array with many more beams, which would transmit a higher overall beam current, depending quadratically on the array size. 

In previous work\cite{Persaud}, we demonstrated acceleration of ion beams through four acceleration gaps and quadrupole focusing through a pair of electrostatic quadrupoles \cite{Persaud2}.   In this work, we began by installing only the matching section downstream of the injector.  The filament driven RF $Ar^+$ source with a $3\times3$ array of extraction electrodes injected ions to the matching section\cite{Ji2016}.  The exiting beam was diagnosed with an $Al_2O_3$ scintillator and image-intensified CCD camera.  Following these observations, the RF accelerator section was added downstream of the matching section, with four acceleration gaps and two quadrupoles.  The quadrupole and RF acceleration electrodes are fabricated from laser-cut copper clad PC-boards.  The radius of the quadrupole and RF electrode apertures are 1~mm, half the matching section aperture.  The exiting total ion beam current was measured in a Faraday cup, as shown in Fig.~\ref{fig:setup}.  

\section{Results and Discussion}

\begin{figure}[ht!]
  \centering
  \includegraphics[width=0.95\linewidth]{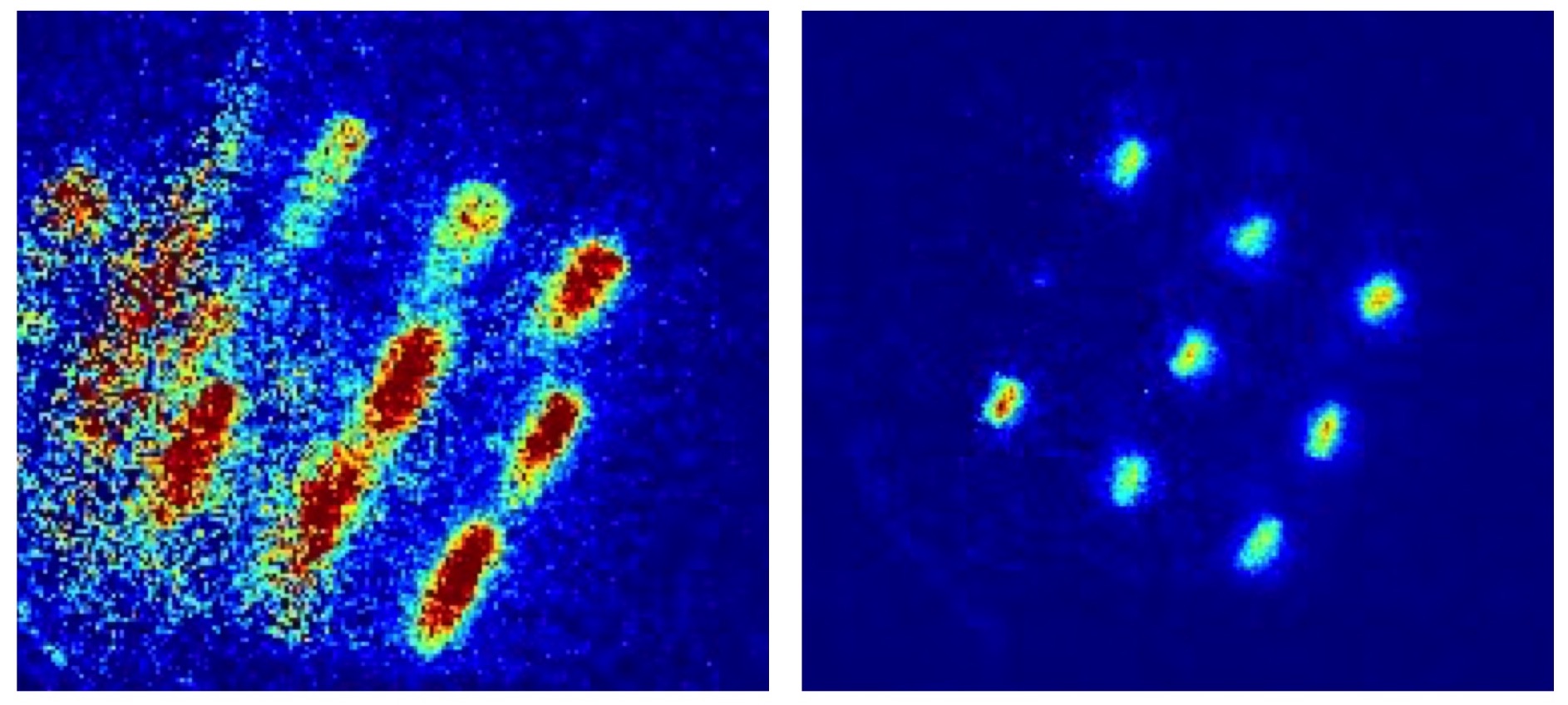}
  \caption{(Color online) Left: Scintillator beam image for only the first matching quadrupole energized. Right: When all quadrupoles are energized each beam has a FWHM approximately half that of the beam profile when the matching section is off. The light intensity varies among the beams due to background subtraction issues related to significant background light from the ion source filament.}
  \label{fig:scintillator}
\end{figure}

Each quadrupole was first energized separately and the expected ellipitical beam pattern was observed in accordance with the quadrupole polarity setting.  Figure~\ref{fig:scintillator} shows an example for the first matching quadrupole energized.  When all the quadrupoles are energized, the beams were more tightly-focused (the FWHM reduced by a factor 0.5).

\begin{figure}[ht!]
  \centering
  \includegraphics[width=0.95\linewidth]{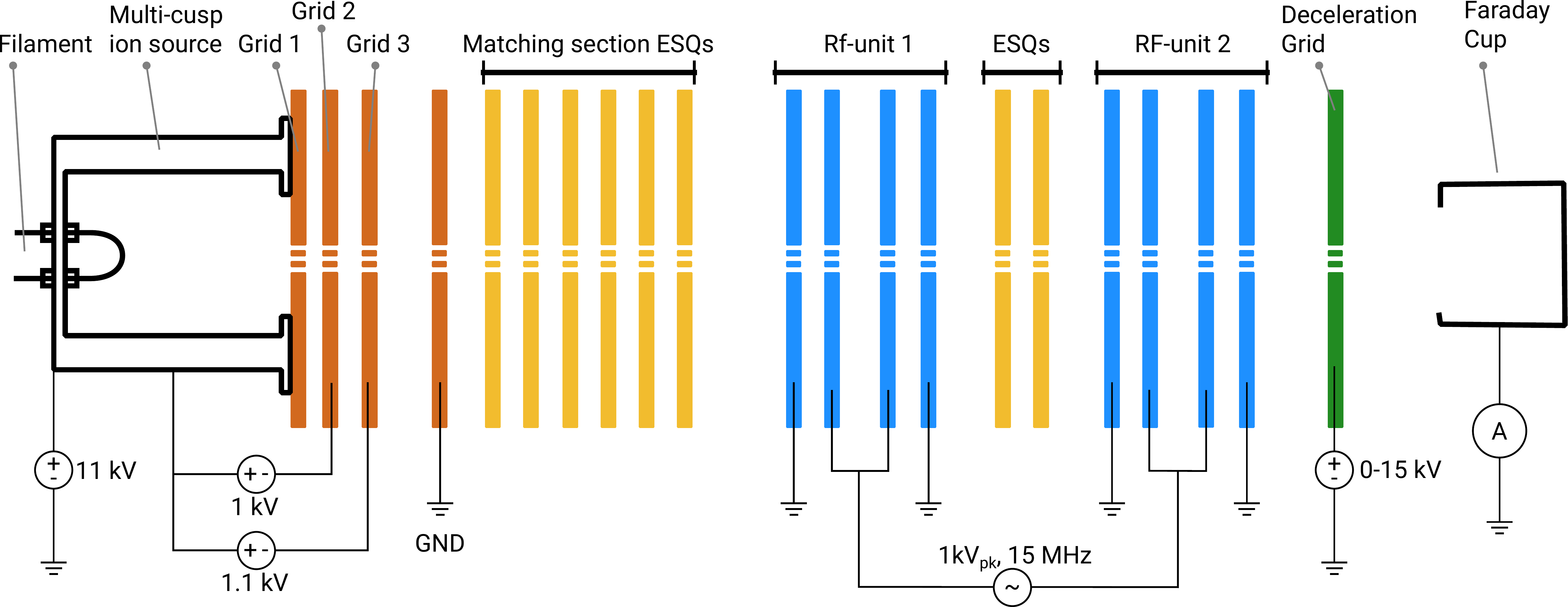}
  \caption{(Color online) Schematic of the experiment.  An argon plasma is generated in the multi-cusp ion source which floats at 11 kV. Ions are extracted through an aligned $3 \times 3$ hole plate assembly (grids) to ground potential and injected into the matching quadrupoles. For the first measurements the ion beam distribution was characterized by a scintillator after the matching section.  In subsequent measurements with an integrated system, the accelerator section (four RF acceleration gaps and two quadrupoles) were added as shown.  A Faraday cup detected the transmitted ion current.}
  \label{fig:setup}
\end{figure}

\begin{figure}[ht!]
  \centering
  \includegraphics[width=0.95\linewidth]{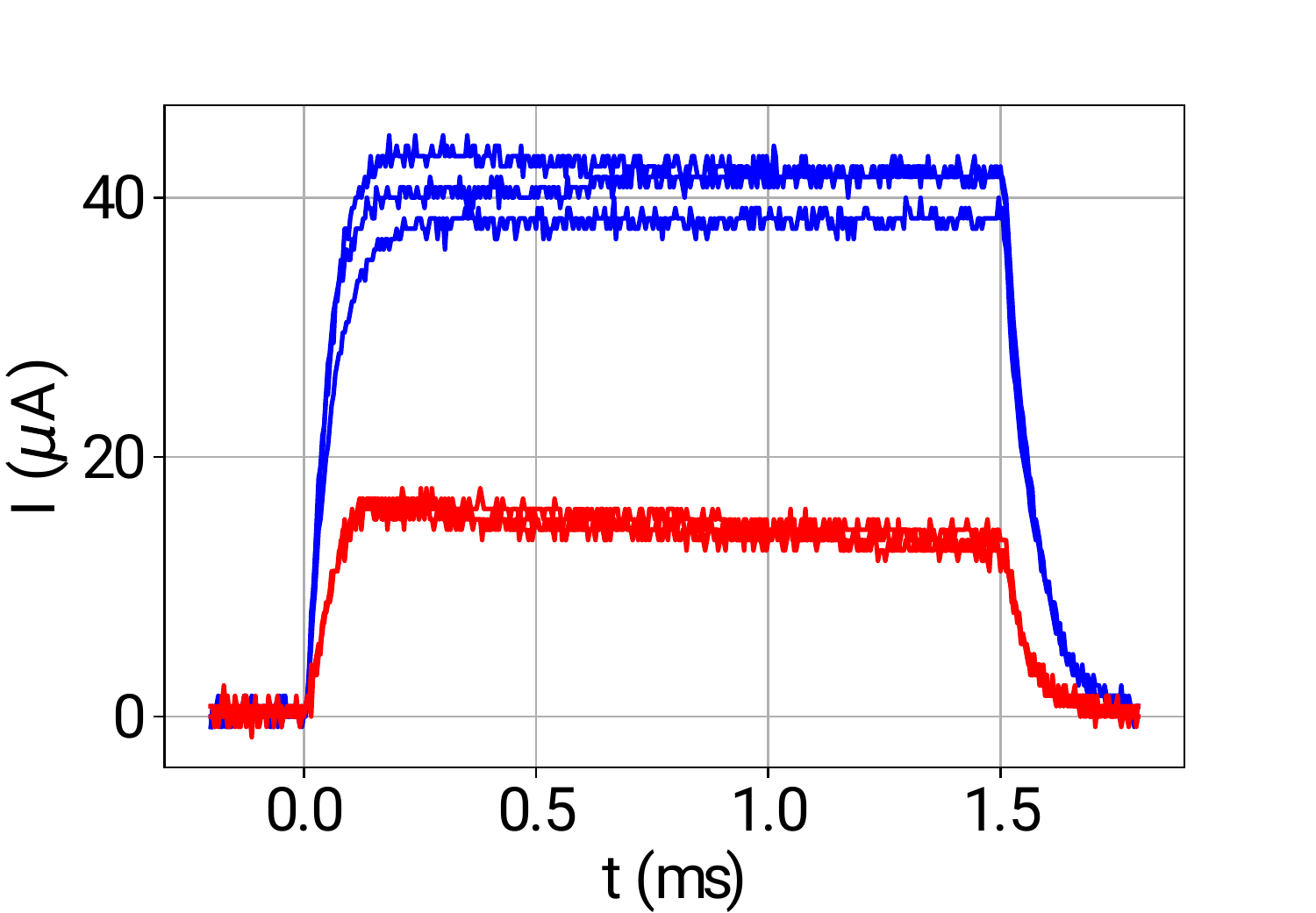}
  \caption{(Color online) The beam current detected by the Faraday cup downstream of the accelerator section shows approximately $2\times$ more current with the matching section on (blue) vs off (red). Three ion pulses are shown for each case.}
  \label{fig:faraday-cup}
\end{figure}

The ion source was pulsed for 1.5 ms, generating a nearly constant current during the pulse.  Figure~\ref{fig:faraday-cup} shows that the transmitted beam current was approximately $2\times$ greater with the matching section energized.

We demonstrated the operation of the integrated assembly, including RF acceleration and quadrupole focusing in the acceleration section as shown in Fig.~\ref{fig:setup}.
The RF acceleration electrodes were driven from a common RF amplifier.  Thus, the spacing between spacing between successive acceleration gaps was set to $\beta \lambda / 2$, as is common in many RF accelerators, where $\beta = v/c$ and $\lambda = c / \nu$,  the RF wavelength. Thus particles near the peak of the acceleration waveform are subject to the same acceleration field at each gap.

The retarding potential of the decelerating grid was varied within the range 
$8 \le V_g \le 14 \: kV$, and the injected ion energy was $E_i = 11$ keV.  Since the RF frequency was 13.5~MHz, the relatively long beam pulses span many RF oscillations. Consequently, ions out of phase with the peak accelerating part of the RF waveform received less acceleration or were decelerated, and a continuum of ion energies is produced between $E_i - \delta E$ and $E_i + \delta E$.  Argon ion energies up to ~13~keV were detected through the four acceleration gaps for 1.5~ms duration pulses. Extrapolating to many more acceleration gaps,  the slower ions would be further out of phase and eventually lost, and the higher energy ions would be preferentially transported and accelerated. 

Some voltage-breakdown in the matching section was observed.  Contributing factors are secondary-electrons from ion lost to the electrodes.  This will be remedied by tuning the envelope solution.  Also, the matching section has a rather closed geometry, and we will add side-vents with higher pumping conductance to achieve a lower pressure within the matching section.

\section{Conclusion and Outlook}

We have fabricated a compact, multiple-beam quadrupole matching system to transform the round beams exiting a multiple beam ion-source to the parameters required for high transmission of ions in an alternating gradient focusing system of an RF accelerator.  The electrode structure substrates are 3D-printed monomer resins with nickel coating to define the quadrupole electrostatic fields, and parallel beams have a center-to-center spacing of 5 mm.  The electrode lengths are 2~mm with a 0.5 mm gap between successive quadrupoles in the $z$ direction.  

Scintillator measurements of the beam exiting the structure show the expected quadrupole focusing effect. When all the quadrupoles are energized, the focused intensity is increased and the beam diameter reduced. Faraday cup measurements also show that the transmitted current is more than doubled by the matching section.  Experiments with the matching section coupled to the compact RF accelerator structure with four acceleration gaps including two focusing quadrupoles between acceleration stages showed $Ar^+$ ion acceleration from the additive acceleration from the four gaps.  

The experience gained here point to improvements to the matching section design. Avoiding the warping caused by the nickel coating of the monomer electrode substrate may be accomplished by fabricating the part wholly out of metal in a 3D-printing process. Also, ports or vents on the side of the matching section will lower the local pressure in the path of the beam near the ion source, improving the voltage holdoff. 

We are presently exploring methods for generating several kV of acceleration per acceleration gap with compact RF power sources.  This is important to enable a compact accelerator ($<1$ m long) structure capable of generating 300 keV ions. 

Higher average current and beam-power can be generated by introducing a bunching section \cite{Wangler2008} after the ion source and matching section to more efficiently capture ($ \approx~50\%$) the injected beam to the stable phase part of the RF acceleration waveform. 

Scaling to many more parallel beams and higher current per beam will enable the development of this accelerator technology for high beam power  applications. Neither require significant increases in component cost nor the development of advanced ion sources.
For example, for the 5-mm pitch between focusing channels tested here, there is ample space to increase the number of parallel beams to $\approx 160$ on each 100-mm diameter wafer.  Each beam could have a peak current of 120 $\mu$A, consistent with established current densities from plasma based ion sources (ca. 15 mA/cm$^2$). A total beam current of $>$9 mA can be produced in a compact structure, for a total beam power in the kilowatt range.

\begin{acknowledgments}
We thank Takeshi Katayanagi for excellent technical support.  We are grateful for insightful discussions with Andris Faltens and James Galvin.

This work was supported by the
US Department of Energy through the ARPA-E ALPHA
program under contract DE-AC0205CH11231.
\end{acknowledgments}

\bibliography{matching-paper}

\end{document}